\documentclass[aps,pra,twocolumn]{revtex4-2}

\usepackage{amsmath}
\usepackage{amsthm}
\usepackage{amsfonts}
\usepackage{amssymb}
\usepackage{xcolor,graphicx}
\usepackage{tikz}
\usepackage{color}
\usepackage[pdftex]{hyperref} %para ligar las referencias
\usepackage{dsfont}
\usepackage{bbold}
\usepackage{physics}

\begin{document}
	\title{
		Approximate cubic phase states in a trapped ion
	}
	%----------------------------------
	%----------------------------------
	\author{C. Ventura-Vel\'{a}zquez}
	\email{cventura@ifuap.buap.mx}
	\affiliation{Instituto de F\'{i}sica, Benem\'{e}rita Universidad Aut\'{o}noma de Puebla, 72570 Puebla, Mexico}
	\author{Juan Mauricio Torres}
    \affiliation{Instituto de F\'{i}sica, Benem\'{e}rita Universidad Aut\'{o}noma de Puebla, 72570 Puebla, Mexico}

	\begin{abstract}
        Universal quantum computation with continuous variables cannot be attained solely with a set of Gaussian operations, it requires the addition of a non-Gaussian element, at least of third order in the quadrature operators, such as the cubic phase state. In this work, we present a method to generate a quantum state in the vibrational mode of a trapped ion that exhibits characteristics compatible with the cubic phase state, such as the distinctive oscillating pattern in its Wigner function. This state emerges from the nonlinear Jaynes-Cummings interaction native to the trapped-ion model, and under the assumption of an initial coherent vibrational state with a large occupation number. Consequently, the evolved vibrational state approximates the cubic phase state with high fidelity, and we use the variance of a nonlinear combination of the quadratures to characterize it. Finally, we provide an analytical expression for its cubicity that shows the high performance of the approximate vibrational cubic phase state.
    \end{abstract}
	
	\maketitle
	
	%%%%%%%%%%%%%%%%%%%%%%%%%%%%%%%%%%%%%%%%%%%%%%%%%%%%%%%%%%%%%%%%%%%
	\section{Introduction}
	%%%%%%%%%%%%%%%%%%%%%%%%%%%%%%%%%%%%%%%%%%%%%%%%%%%%%%%%%%%%%%%%%%%
	
	Classical computation relies on the binary system of 0 and 1 to perform calculations to solve problems, while quantum computation takes advantage of the superposition  and entanglement  of states to run algorithms that can be faster for certain tasks, such as factorization of large numbers, simulating quantum systems, and some optimization problems \cite{Nielsen2000}. There are two ways to encode quantum information: discrete and continuous variable (CV) approaches. In the former one, the information is stored with a finite set of distinguishable basis states, for example, the excited and ground states corresponding to the electronic part of a trapped ion \cite{Haeffner2008}. In contrast, the CV approach encodes information in quantum systems whose observables have continuous spectra, such as the position and momentum quadratures of an optical field \cite{Lloyd1999}. A few examples of experimental and theoretical setups for CV quantum computation are light modes \cite{Asavanant2019,ShotaYokoyama2013}, a microwave cavity \cite{CampagneIbarcq2020}, superconducting resonators \cite{Hillmann2020,Kudra2022,AxelM.Eriksson2024}, and optomechanical systems \cite{Rakhubovsky2021,Houhou2022}.
	
	Analogously to their classical counterparts, quantum gates are needed to implement operations on a quantum state. The CV models often use operations such as rotations, displacements, squeezing, and beam splitters, which correspond to a set of gates known as Gaussian \cite{Wu1986,Loudon1987,Sefi2011,Weedbrook2012}. However, quantum computation using only Gaussian gates can be efficiently simulated on a classical computer \cite{Bartlett2002}. Therefore, a non-Gaussian gate is required to go beyond Gaussian dynamics and to have the ability to approximate, with arbitrary accuracy, any unitary transformation \cite{Ghose2007,Sefi2011}. For this reason, the implementation of non-Gaussian resources is often viewed as a key requirement for universal CV quantum computation \cite{Lloyd1999}. Among the candidates for the non-Gaussian element, the canonical choice is the so-called cubic phase gate and its associated cubic phase state (CPS) obtained by applying a nonlinear operation of third order to an infinite squeezed state \cite{Gottesman2001,Ghose2007,Budinger2024}.

    The ideal CPS requires infinite squeezing which is physically impossible in experiments. Therefore, several approximate methods have been theoretically proposed \cite{Hillmann2020,Zheng2021,Konno2021,RieraCampeny2024}, but few experimental setups have been tested \cite{Kudra2022,AxelM.Eriksson2024}. Thus, a reliable way to characterize the non-Gaussian features of the resulting quantum states across diverse approaches is needed. Two widely used methods to compare quantum states are the state fidelity and the Wigner function. However, the former is insufficient, as it can assign high values to quantum states that are a poor approximation to a CPS \cite{Konno2021}. For the latter, several measurements are needed to reconstruct a Wigner function, and its negative parts are sensitive to noise or losses in the studied system \cite{Agarwal2007}. One can find a better alternative by noting that the cubic phase gate causes nonlinear effects on the quadratures of the field \cite{PMarek2011,Konno2021} and, as a result, a nonlinear squeezing (NLS) is induced. This NLS acts as a more precise method to characterize the non-Gaussian features of the studied state that are compatible with a CPS \cite{Darren2022}. The advantages of the NLS are its robustness against noise and that it requires few measurements to estimate it \cite{Moore_2019}.
    
	In this work, using the nonlinear interaction between the electronic and the vibrational degrees of freedom of a single trapped ion, we propose the generation of an approximate CPS in the vibrational mode of the ion. We rely on the approximate separable solution to the time-evolution given in Ref.~\cite{MauricioTorres2025}, allowing us to study non-Gaussian properties of the motion. We find that the Wigner function of the vibrational mode presents characteristics of a CPS, and that a high fidelity between them is achieved. Furthermore, the NLS of our generated vibrational state shows that it is suitable for use as an ancillary state for the cubic phase gate in CV clusters \cite{Gu2009}. Finally, we obtain an analytical expression for the cubicity in the limit of a large initial occupation number.
	
	The manuscript is organized as follows. In Sec. \ref{sec:ion}, we introduce the trapped-ion model and present an approximate solution for the time-dependent state describing a separable evolution between electronic and vibrational degrees of freedom. We test its validity by calculating the fidelity with respect to the numerically exact solution. In Sec. \ref{sec:CPS}, we present the method to generate approximate CPS in the vibrational part of the ion, and we give an analytical expression for its cubicity. To evaluate the generated state, we calculate its fidelity with respect to an ideal CPS, and with the aid of the Wigner function, we visualize its behavior in phase-space. In Sec. \ref{sec:NLS}, we introduce the NLS as the tool to evaluate the non-Gaussian features of the vibrational state that make it a good approximation to the CPS. Additionally, we use the NLS to study the cubicity, and we compare it with the values obtained using the analytical expression in the previous section. Finally, we present our conclusions and remarks in Sec. \ref{sec:conclusions}.
    
    %%%%%%%%%%%%%%%%%%%%%%%%%%%%%%%%%%%%%%%%%%%%%%%%%%%%%%%%%%%%%%%%%%%
	\section{The model} \label{sec:ion}
	%%%%%%%%%%%%%%%%%%%%%%%%%%%%%%%%%%%%%%%%%%%%%%%%%%%%%%%%%%%%%%%%%%%
	
	In this section, we introduce the model and develop the analytical solution for the time-evolution of the quantum state of a trapped ion \cite{MauricioTorres2025}. Under the assumption of a large initial coherent state for the vibrational mode, we are able to give an approximate separable solution of the time-dependent state vector. We validate  this approximation by evaluating the fidelity with respect to the numerically exact evolved state.
	
	%%%%%%%%%%%%%%%%%%%%%%%%%%%%%%%%%%%%%%%%%%%%%%%%%%%%%%%%%%%%%%%%%%%
	\subsection{Nonlinear Jaynes-Cummings model}
	%%%%%%%%%%%%%%%%%%%%%%%%%%%%%%%%%%%%%%%%%%%%%%%%%%%%%%%%%%%%%%%%%%%
	
	We consider the nonlinear Jaynes-Cummings model for a single trapped ion \cite{Vogel1995}. In the first vibrational sideband, we can write the interaction potential as
	\begin{eqnarray}
    \label{eq:V}
		\hat{V}
		&=&
		\hbar\,\Omega\left[
		f(\hat{a}^{\dag} \hat{a})\, \hat{a}\, 
        \hat{\sigma}_{+}
		+
		\hat{a}^{\dag} f(\hat{a}^{\dag} \hat{a})\, 
        \hat{\sigma}_{-}
		\right],
	\end{eqnarray}
	where $\hbar$ is the reduced Planck constant, $\Omega$ is the frequency of exchange energy between the electronic and vibrational states, $\hat{\sigma}_{\pm} = (\hat{\sigma}_{x} \pm i\, \hat{\sigma}_{y})/2$ describes the electronic state of the trapped ion using the Pauli matrices, $\hat{a}$ ($\hat{a}^{\dag}$) is the bosonic operator of annihilation (creation) with commutaion relation $[\hat{a},\, \hat{a}^{\dag}] = 1$, and $\eta$ is the Lamb-Dicke parameter. The function $f(\hat{a}^{\dag} \hat{a})$ has the following intensity-dependent form
	\begin{eqnarray}
    \label{eq:f_n}
		f(\hat{n})
		=
		\frac{\eta\, e^{-\eta^{2}/2}}{\hat{n}+1}\,
		L_{\hat{n}}^{(1)} (\eta^{2}), \quad \hat{n} = \hat{a}^{\dag} \hat{a},
	\end{eqnarray}
	where $\hat{n} $ is the number operator and we use the definition of the associated Laguerre polynomials \cite{WeberArf2003}.
	
	The interaction potential commutes with the constant of motion $\hat{I} = \hat{n} +\hat{\sigma}_{+}\hat{\sigma}_{-}$ which has the set of eigenstates 
    $\left\lbrace  \ket{e,\, n-1},\, \ket{g,\, n} \right\rbrace $ for $n >0$ and the singlet $\ket{g,\, 0}$. Using them, we can write the eigenstates and eigenenergies for $\hat{V}$ as
    \begin{subequations}
        \begin{eqnarray}
        \ket{ E_{n}^{\pm} }
        &=&
        \dfrac{
            \ket{ e,\, n-1 }
            \pm
            \ket{ g,\, n }
        }{\sqrt{2}},
        \quad
        n \geq 1,
        \\
        E_{n}^{\pm}
        &=&\pm\hbar\,\Omega_n,
        \quad 
        \Omega_{n}
        =
        \Omega
        \bra{ n }
            \hat{a}^{\dag}\, f(\hat{n})
        \ket{ n-1 },
        \label{eq:energy}
    \end{eqnarray}
    \end{subequations}
    where the eigenenergies are expressed in terms of the eigenfrequencies $\Omega_n$. For $n = 0$, the singlet $\ket{g,\, 0}$ is an eigenstate with vanishing energy $E_{0} = 0$. 
    
    %%%%%%%%%%%%%%%%%%%%%%%%%%%%%%%%%%%%%%%%%%%%%%%%%%%%%%%%%%%%%%%%%%%
    \subsection{Time-dependent state}
    %%%%%%%%%%%%%%%%%%%%%%%%%%%%%%%%%%%%%%%%%%%%%%%%%%%%%%%%%%%%%%%%%%%
    
    Having solved the eigenvalue problem, it is possible to write the exact time-evolution for any initial state of the trapped ion as in \cite{MauricioTorres2025}. Here, however, we will only work with the specific initial condition where the electronic part is in a symmetric or antisymmetric superposition of the ground and excited states, while the vibrational part is prepared as a coherent state $\ket{\alpha}$, with real-valued amplitude $\alpha$. The initial product state can be expressed in the following way
    \begin{subequations}
        \begin{align}
            \ket{ \Psi (0) }
            &=
            \dfrac{\ket{e} \pm \ket{g}}
            {\sqrt{2}} 
            \otimes
            \ket{ \alpha },
            \label{eq:psi0}
            \\
            \ket{ \alpha }
            &=
            \sum_{n=0}^{\infty}
                p_{n}\, 
                \ket{n},
            \quad
            p_{n}
            =
            e^{-|\alpha|^{2}/2}\,
            \alpha^{n} /\sqrt{n!}.
       \end{align}
    \end{subequations}
    Assuming that the initial vibrational state has a large occupation number, that is $|\alpha|^{2} = N \gg 1$, it is possible to consider the relation $p_{n} \approx p_{n \pm 1}$ that leads to the approximate initial state $\ket{ \Psi (0) } \approx \sum_{n} p_{n}\, \ket{ E_{n}^{\pm} }$. Consequently, the time-dependent state, given by $\ket{ \Psi (t) } = e^{-i\, \hat{V}\,t/\hbar}\, \ket{ \Psi (0) }$, can be written in an approximate fashion as
    \begin{eqnarray}
        \ket{ \Psi (t) }
        &\approx&
        \sum_{n=0}^{\infty}
            p_{n}\, e^{\mp i\, \Omega_n\, t}\,
            \ket{ E_{n}^{\pm} },
        \nonumber
        \\
        &\approx&
        \sum_{n=0}^{\infty}
            \dfrac{
                e^{\mp i\, \varphi_{n}\, t}\,
                \ket{e}
                \pm
                \ket{g}
            }
            {\sqrt{2}}
            \otimes
            e^{\mp i\, \Omega_n\, t}\, p_{n}\,
            \ket{n},
            \label{eq:PsiSolution0}
    \end{eqnarray}
    where $\varphi_{n} = \Omega_{n+1} -\Omega_{n}$. Note that this solution is independent of the form of the eigenfrequencies $\Omega_{n}$. The generalization to complex-valued $\alpha$ can be obtained as in \cite{MauricioTorres2025} with a unitary transformation in terms of the constant of motion $\hat{I}$. In particular, the result is valid for the Jaynes-Cummings model, with $f(\hat{n})=1$, where this type of solutions is well known \cite{Gea-Banacloche1990,Jarvis2009}. For that case, we get $\Omega_n=\Omega\sqrt{n}$, which makes the phase $\varphi_{n}$ approximately constant for large $n$, leading to an independent vibrational dynamics in Eq. \eqref{eq:PsiSolution0}.

    %%%%%%%%%%%%%%%%%%%%%%%%%%%%%%%%%%%%%%%%%%%%%%%%%%%%%%%%%%%%%%%%%%%
    \subsection{Separable solution with a non-Gaussian state}
    %%%%%%%%%%%%%%%%%%%%%%%%%%%%%%%%%%%%%%%%%%%%%%%%%%%%%%%%%%%%%%%%%%%
    
    In our previous work \cite{MauricioTorres2025}, we found conditions under which $\Omega_{n}$ is approximately linear for very large $n$, leading to a constant $\varphi_{n}$ in Eq. \eqref{eq:PsiSolution0}. Here, we focus on the same conditions with moderate values of $n$ where a correction to the linear behavior of $\Omega_{n}$ is observed, but preserving the constant value of $\varphi_{n}$. This third-order correction to the eigenfrequencies allows us to describe the generation of vibrational non-Gaussian states.
     
    According to Eqs. \eqref{eq:f_n} and \eqref{eq:energy}, the explicit form of the eigenenergies is given by the function
    \begin{eqnarray}
		\Omega_{n}
		&=& 
        \Omega\, 
        \frac{\eta\, e^{-\eta^{2}/2}}{\sqrt{n}}\,
		L_{n-1}^{(1)} (\eta^{2})
		\approx 
        \Omega\,
		J_{1} (2\, \eta\, \sqrt{n}),
	\end{eqnarray}
   where we used an approximation in terms of the Bessel function of the first kind which is valid for small values of $\eta$ \cite{Szego1975}.
   
   %.
	\begin{figure}[tbp!]
		\centering
		\includegraphics[width=0.99\linewidth]{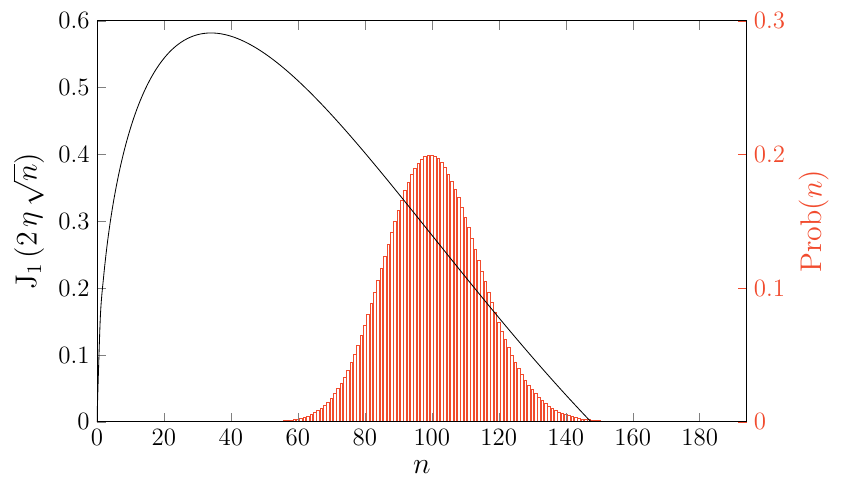}
		\caption{	\label{fig:effective_H}
			The black curve illustrates the Bessel function of the first kind, $J_{1}$ as a function of $n$, and the red bars show the probability distribution of the initial coherent vibrational state, $\ket{\alpha}$, around the point $N = |\alpha|^{2} = 100$, which corresponds to an inflection point of the Bessel function. We used a value of $\eta \approx 0.15773082$ such that the condition $ (2\,\eta)^{2} N = 9.95160559 $ is met.
		}
	\end{figure}
	%.
    
    In Fig. \ref{fig:effective_H}, we plot the behavior of $\Omega_{n} /\Omega$ as a function of the number $n$ with the overlapped Poisson distribution of the initial coherent state with an occupation number $N = |\alpha|^{2} = 10^{2} \gg 1$ and the Lamb-Dicke parameter $\eta \approx 0.15773082$. Due to the spread of the probability distribution of the initial coherent state around $n = N$, the dynamics can be approximated using the Taylor expansion of $\Omega_{n}$ up to the third power of $n$, namely
    \begin{eqnarray}
    \label{eq:Omega_Taylor}
		\Omega_{n} 
		&\approx&
		\omega_{0}
		+
		\omega_{1} \left(
			     n - N
		\right)
		+
		\omega_{3} \left(
			     n - N
		\right)^3,
	\end{eqnarray}
    \begin{equation}
    \label{eq:omegaj}
			\omega_{j}
			=
            \frac{1}{j!}
			\left.
            \frac{d^{j} \Omega_{n}}
            {dn ^{j}}\right|_{n=N},
            \quad 
            N = \frac{y_1}{(2\eta)^{2}},
            \quad
            y_{1} = 9.9516,
    \end{equation}
    which can be computed in terms of the Bessel function. We choose $N$ and $\eta$ such that the second derivative vanishes, i.e., $\omega_{2} = 0$. This occurs when the following condition is fulfilled $(2\,\eta)^{2} N = y_{1} = 9.95160559$, expressed also in Eq. \eqref{eq:omegaj}, as explained in \cite{MauricioTorres2025}. In this case, one can verify the approximate constant behavior $\varphi_{n} \approx \omega_{1} +\omega_{3}$. Using the initial state in Eq. \eqref{eq:psi0} with the plus sign, it is now possible to arrive at a separable time-dependent state in terms of a unitary transformation acting solely on the vibrational part in terms of the Hermitian operator $\Omega_{\hat{n}}$, obtained by replacing $n\to\hat n$ in Eq. \eqref{eq:Omega_Taylor}, as
    \begin{eqnarray}
    \label{eq:evol_sep}
        \ket{ \Psi_{s}(t) }
		&\approx&
		\dfrac{e^{- i\, (\omega_{1}+\omega_{3})\, t} \ket{e} + \ket{g}}
		{\sqrt{2}}
		\otimes
        \ket{\psi(t)},
    \end{eqnarray}
    with the time-dependent vibrational state 
    \begin{equation}
    \label{eq:psi_vib}
        \ket{\psi(t)}
        =
        \hat{U} (t)
        \ket{\alpha},
        \quad 
        \hat{U}(t)
        =
        e^{- i\, \Omega_{\hat{n}}\, t}.
    \end{equation}
    Note that this solution is valid when the Poisson distribution of the coherent state is centred around the mean occupation number where the second derivative of $\Omega_{n}$ vanishes. We have previously explored these solutions \cite{MauricioTorres2025}, as it allows for a dynamical description in terms of coherent states for very large values of $N$. Here, we choose moderately large values of $N$, where the third-order contribution in Eq. \eqref{eq:Omega_Taylor} is non-negligible, leading to a nonlinear evolution capable of generating an approximate CPS as will be shown in the next section. 

    More explicitly, the time-dependent vibrational state of the trapped ion can be expressed as a nonlinear unitary transformation applied to a complex-valued coherent state, that is
    \begin{equation}
    \label{eq:psi_vib2}
        \ket{\psi(t)}
        =
        e^{i\,\delta\,t }\,
        e^{-i\omega_3(\hat n-N)^3t}
        \ket{e^{-i\omega_1 t}\alpha},
    \end{equation}
    with $\delta = \omega_{1}\, N - \omega_{0}$. For negligible values of $\omega_{3}$, this corresponds to a time-dependent coherent state, whose amplitude attains real values at integer multiples of the time $\pi / |\omega_{1}|$. The first time this happens, apart from zero, is $\tau = \pi / |\omega_{1}|$ and it is known as the revival time for general electronic states, because the two field components meet at opposite sides in phase-space originating the revivals of Rabi oscillations in the system observables \cite{Jarvis2009,Gea-Banacloche1990,MauricioTorres2025}. We choose the revival time to study the evolved vibrational state and we will show in Sec. \ref{sec:CPS} that an approximate CPS can be generated.
    
    %.
	\begin{figure}[tbp!]
		\centering
		\includegraphics[width=0.95\linewidth]{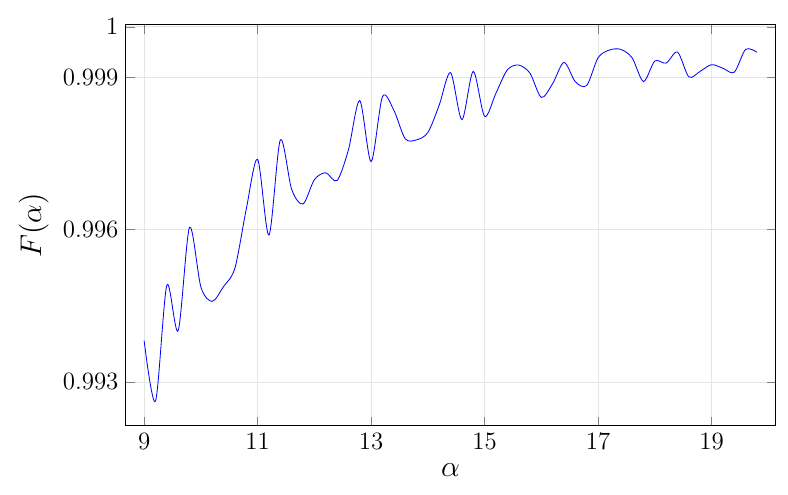}
		\caption{	\label{fig:fid_full}
			Fidelity between the exact numerical evolution and the approximate separable solution, in Eq. \eqref{eq:evol_sep}, for different values of the initial coherent amplitude $\alpha$. We fixed the time at $\tau = \pi/|\omega_{1}|$.
		}
	\end{figure}
	%.
    
    Before continuing, let us validate the accuracy of our approximate solution. For this, we evaluate the fidelity between the numerically exact evolved state with the interaction potential $\hat{V}$ and the separable state given by Eq. \eqref{eq:evol_sep} as a function of the initial coherent amplitude $\alpha$:
	\begin{eqnarray}
		F(\alpha)
		&=&
		\left| 
		      \bra{ \Psi_{s} (\tau) }\,
		      e^{-i\, \hat{V}\, \tau}\,
		      \ket{ \Psi (0) }  
		\right| ^{2}.
	\end{eqnarray}
    The result is displayed in Fig. \ref{fig:fid_full}, and we notice that the fidelity gets closer to one as the initial coherent amplitude, $\alpha$, increases, which confirms the validity of our separable solution given in Eq. \eqref{eq:evol_sep}. In particular, for $\alpha \geq 11$ we have a fidelity greater than $0.996$.
	
	%%%%%%%%%%%%%%%%%%%%%%%%%%%%%%%%%%%%%%%%%%%%%%%%%%%%%%%%%%%%%%%%%%%
	\section{Vibrational cubic phase state} \label{sec:CPS}
	%%%%%%%%%%%%%%%%%%%%%%%%%%%%%%%%%%%%%%%%%%%%%%%%%%%%%%%%%%%%%%%%%%%

    %.
	\begin{figure}[tbp!]
		\centering
		\includegraphics[width=0.95\linewidth]{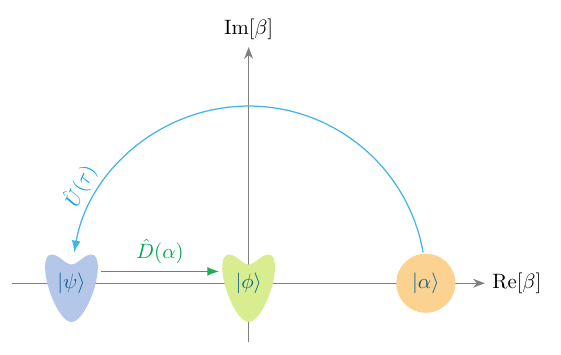}
		\caption{	\label{fig:evol_cohe}
        Sketch of the path followed by the initial coherent vibrational state $\ket{\alpha}$ in the phase-space. According to the evolution operator $\hat{U}$, at time $\tau = \pi/|\omega_{1}|$, the final state is located at the opposite side, and due to the nonlinearity of $\Omega_{\hat{n}}$, it is deformed in such a way that its characteristics are similar to a CPS. The purpose of the displacement is to relocate it to the origin of the phase space and characterize it with the NLS.
		}
	\end{figure}
	%.

	In this section and onward, we focus our attention only on the vibrational state as, previously explained, it is possible to approximate the time-dependent state of the trapped ion as two decoupled dynamics: one for its electronic part and the other for its vibrational mode. Here, we will show that the vibrational state can be tailored to an approximate vibrational CPS, which is defined as \cite{Gu2009}:
	\begin{eqnarray}
    \label{eq:CPS}
		\ket{\gamma}
		=
		e^{\, i\,\gamma\,\hat{x}^{3}}\,
		\ket{0},
        \quad
        \hat{x} 
        =
        \dfrac{\hat{a}^{\dag} +\hat{a}}{\sqrt{2}},
	\end{eqnarray}
	where $\hat{x}$ is the dimensionless position operator and $\gamma$ the so-called cubicity. Furthermore, we will test the  validity of our generated state through the fidelity with the ideal one, and compare its behavior in phase-space with the aid of the Wigner function.

    %%%%%%%%%%%%%%%%%%%%%%%%%%%%%%%%%%%%%%%%%%%%%%%%%%%%%%%%%%%%%%%%%%%
    \subsection{Generation of the cubic phase state}
    %%%%%%%%%%%%%%%%%%%%%%%%%%%%%%%%%%%%%%%%%%%%%%%%%%%%%%%%%%%%%%%%%%%

    %.
	\begin{figure}[t]
		\centering
		\includegraphics[width=0.95\linewidth]{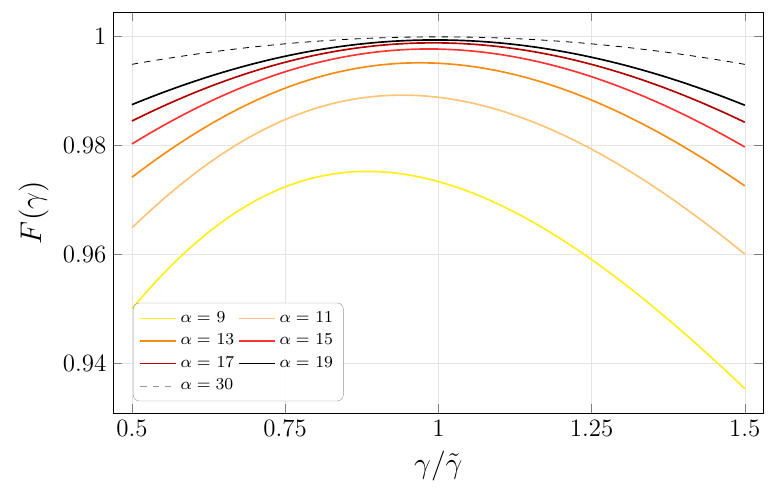}
		\caption{	\label{fig:fid_gam}
        The fidelity between the vibrational state obtained with the separable state in the displaced frame, Eq. \eqref{eq:psi_dis}, and the ideal cubic phase state, Eq. \eqref{eq:CPS}. Each curve corresponds to a different value of the initial coherent amplitude, $\alpha$. Additionally, we rescaled the cubicity parameter, $\gamma$, with the correspondent $\tilde{\gamma}$, Eq. \eqref{eq:gamma}, for each $\alpha$.
		}
	\end{figure}
	%.

    %.
	\begin{figure*}[t!]
		\centering
		\includegraphics[width=0.85\linewidth]{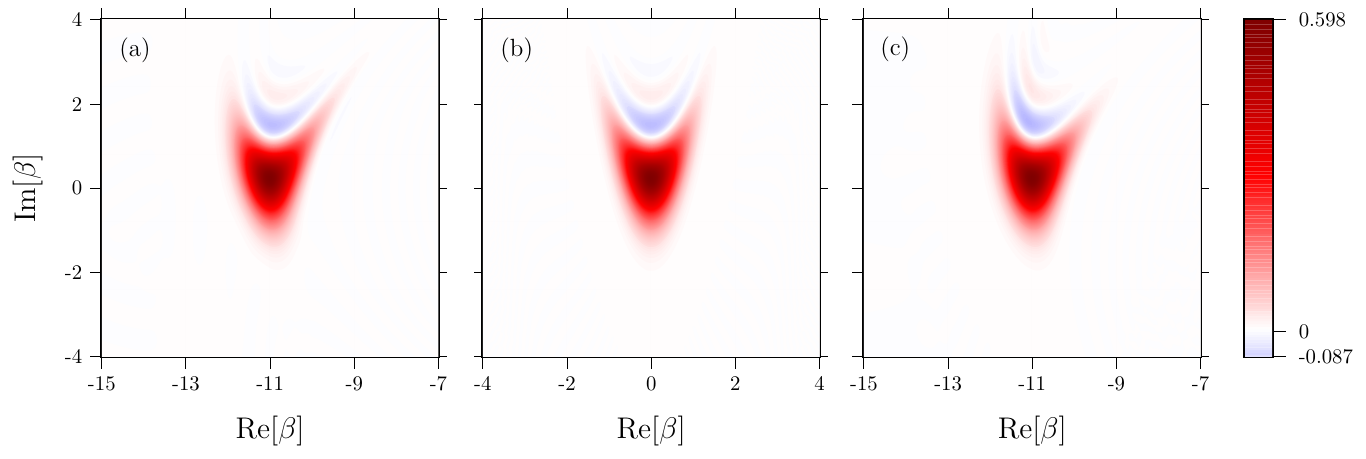}
		\caption{	\label{fig:wigner}
			The Wigner function of the vibrational state for: (a) the exact time-evolution generated by the interaction potential in Eq. \eqref{eq:V}, (b) the cubic phase state with cubicity $\tilde{\gamma} = 0.2862$ given by Eq. \eqref{eq:gamma}, and (c) the time-evolution given by the approximate separable state, Eq. \eqref{eq:evol_sep}. The initial state is Eq. \eqref{eq:psi0} with the plus sign, the coherent amplitude is $\alpha = 11$ and the time $\tau = \pi/|\omega_{1}|$ for (a) and (c).
		}
	\end{figure*}  
    %.
    
    The vibrational state arising from the interaction and given in Eq. \eqref{eq:psi_vib2} results from a unitary transformation whose generator involves the cubed number operator, i.e., $(\hat n-N)^3$. This transformation can be related to a cubic phase gate in the displaced vibrational picture; then, the position operator emerges as
    \begin{equation}
        \hat{D}(\alpha)\,
        \hat{n}\,
        \hat{D}^{\dag}(\alpha)
        =
        \hat{n}-\sqrt{2}\,\alpha\,\hat{x} +N,
    \end{equation}
    where we introduced the displacement operator for a complex-valued argument $\beta$ as $\hat{D}(\beta) = e^{\beta\,\hat{a}^{\dag} +\beta^{*}\,\hat{a}}$. Noticing that at the revival time $\tau = \pi / |\omega_1|$, the coherent state in Eq. \eqref{eq:psi_vib2} results in $\ket{-\alpha}$, one can consider that the displaced vibrational state is transformed as
		\begin{align}
        \label{eq:psi_dis}
			\ket{\phi}
			&=
			\hat{D}(\alpha)\, 
            \ket{ \psi(\tau) }
			=
			e^{i\, \delta\, \tau}\,
			e^{-i\,\omega_{3}\,\tau(\hat{n} - \sqrt{2}\,\alpha\,\hat{x})^{3}}\,
			\ket{0}.
		\end{align}
    where the time-evolution operator, $\hat{U}(\tau)$, is defined in Eq. \eqref{eq:psi_vib}. Remarkably, the displaced state in Eq. \eqref{eq:psi_dis} is quite similar to a CPS except for the term proportional to the number operator in the exponential. However, in the displaced picture, the initial state is the vacuum; therefore, for an increasing value of the coherent amplitude, $\alpha = \sqrt{N}$, the effect of the contribution due to the $\hat{n}$ diminishes, and one can neglect it provided $\alpha$ is large enough, leading to the following approximation
		\begin{align}    
        \label{eq:mean_field}
			\ket{\phi}
			\approx
			e^{i\, \delta\, \tau}\,
			e^{i\, \tilde{\gamma}\, \hat{x}^{3}}\,
			\ket{0},			
		\end{align}
    where the cubicity for this specific state is expressed as
    \begin{equation}
        \label{eq:gamma}
        \tilde{\gamma}
		=
		\omega_{3}\, \tau\,
		( \sqrt{2}\,\alpha )^{3}
		=
		\dfrac{2^{3/2}\, \pi\, J_{1}^{'''}(\sqrt{y_{1}}) y_1^2 }
        {6\,\alpha |J_{1}^{'}(\sqrt{y_{1}})|}
        \approx
        \frac{3.148803}{\alpha},
    \end{equation}
    and the numerical values for the first and third derivatives of the Bessel function are $J_{1}^{'}(\sqrt{y_{1}}) = -0.06284499$ and $J_{1}^{'''}(\sqrt{y_{1}}) = 0.00134923$, respectively. Thus, in the displaced picture, it is straightforward to see that the evolved vibrational state is an approximation to the CPS with cubicity given by Eq. \eqref{eq:gamma}. Additionally, for very large values of the initial coherent amplitude, we find that the cubicity tends to zero, $\tilde{\gamma} \to 0$, which coincides with the linear evolution reported in \cite{MauricioTorres2025}, in other words, the resulting linear behavior preserves the coherent state.
    
    In Fig. \ref{fig:evol_cohe}, we illustrate the time-evolution of the initial coherent vibrational state in the phase-space representation; according to Eq. \eqref{eq:psi_vib} and after the time $\tau = \pi/|\omega_{1}|$, the resulting state has travelled half a circumference to reach the opposite site with a deformed shape. Also, in the sketch, we show that the evolved vibrational state is translated to the origin of the phase-space with the displacement operator $\hat{D}(\alpha)$, as a result we get a vanishing mean position value, $\langle \hat{x} \rangle = 0$.
    
    %%%%%%%%%%%%%%%%%%%%%%%%%%%%%%%%%%%%%%%%%%%%%%%%%%%%%%%%%%%%%%%%%%%
	\subsection{Fidelity of the vibrational cubic phase state}
    %%%%%%%%%%%%%%%%%%%%%%%%%%%%%%%%%%%%%%%%%%%%%%%%%%%%%%%%%%%%%%%%%%%
    
    Now, we are in a position to study the similarities of our approximate vibrational CPS with the ideal one. To quantify the accuracy of our approximation, we now compute numerically the fidelity between an ideal CPS and the displaced vibrational state given in Eq. \eqref{eq:psi_dis}, this is
    \begin{eqnarray}
		F (\gamma)
		&=&
		\left| 
    		\left\langle \gamma \right.
            \left| \phi \right\rangle 
		\right| ^{2}.
	\end{eqnarray}
	We plot this fidelity in Fig. \ref{fig:fid_gam} as a function of the rescaled cubicity $\gamma/ \tilde{\gamma}$, with $\tilde{\gamma}$ given in Eq. \eqref{eq:gamma}, and for several values of the initial coherent amplitude, $\alpha$. The cubicity value $\gamma$ is varied for the CPS in Eq. \eqref{eq:CPS}, whereas the corresponding $\tilde\gamma$ in our generated state, Eq. \eqref{eq:psi_dis} is fixed by the value of $\alpha$. We note that for larger values of $\alpha$ the fidelity has a maximum closer to $\gamma = \tilde{\gamma}$, which validates the approximation made in Eq. \eqref{eq:mean_field}. In particular, we obtain fidelity values larger than 0.96 even for moderate values of $\alpha \geq 11$.
    
    %%%%%%%%%%%%%%%%%%%%%%%%%%%%%%%%%%%%%%%%%%%%%%%%%%%%%%%%%%%%%%%%%%%
	\subsection{Wigner function}
    %%%%%%%%%%%%%%%%%%%%%%%%%%%%%%%%%%%%%%%%%%%%%%%%%%%%%%%%%%%%%%%%%%%

    Finally, we visualize the non-Gaussian features of the generated vibrational state in phase-space with the aid of the Wigner function defined as
    \begin{eqnarray}
        W(\beta)
        =
        \dfrac{1}{\pi}\,
        \Tr_{v} \lbrace
            \hat{D} (\beta)\,
            e^{i\,\pi\,\hat{n}}\,
            \hat{D}^{\dag} (\beta)\,
            \hat{\rho}_{v}
            \rbrace,
    \end{eqnarray}
    where $\hat{\rho}_{v}$ is the reduced density matrix for the vibrational part, that is obtained by taking the partial trace of the total density matrix over the electronic degrees of freedom, i. e., $\hat{\rho}_{v} = \Tr_{el} \lbrace \hat{\rho} \rbrace$ with $\hat{\rho} = \ket{\Psi(t)} \bra{\Psi(t)}$. This quasi-probability distribution can take negative values, which is a strong signature of the non-classical nature of the studied quantum state. The Wigner function of a CPS shows a unique pattern of oscillations with negative values in the region where the momentum takes positive values \cite{Ghose2007}, that is $\Im[\beta] > 0$, and a quite similar behavior is observed for our generated vibrational state, either with the exact dynamics or our approximate separable solution. We use the open-source software QuTiP \cite{Johansson2012,Johansson2013,Lambert2026} to numerically compute the Wigner functions shown in Fig. \ref{fig:wigner}, which shows a comparison between the ideal CPS and the vibrational state generated using the interaction potential in Eq. \eqref{eq:V} and the vibrational state in Eq. \eqref{eq:psi_vib2}. Remarkably, the oscillating pattern appears clearly in the generated vibrational state, but the global shape seems to be twisted to the right side, this effect is due to the dependence on $\hat{n}$ of the function $f(\hat{n})$ in the interaction potential $\hat{V}$, and consequently in the eigenfrequencies $\Omega_{n}$. Nonetheless, this visual approach using the Wigner function reaffirms the generation of approximate CPS in the vibrational mode of a trapped ion.
    
    %%%%%%%%%%%%%%%%%%%%%%%%%%%%%%%%%%%%%%%%%%%%%%%%%%%%%%%%%%%%%%%%%%%
	\section{Nonlinear squeezing} \label{sec:NLS}
    %%%%%%%%%%%%%%%%%%%%%%%%%%%%%%%%%%%%%%%%%%%%%%%%%%%%%%%%%%%%%%%%%%%
    
	%.
	\begin{figure}[tbp!]
		\centering
		\includegraphics[width=0.95\linewidth]{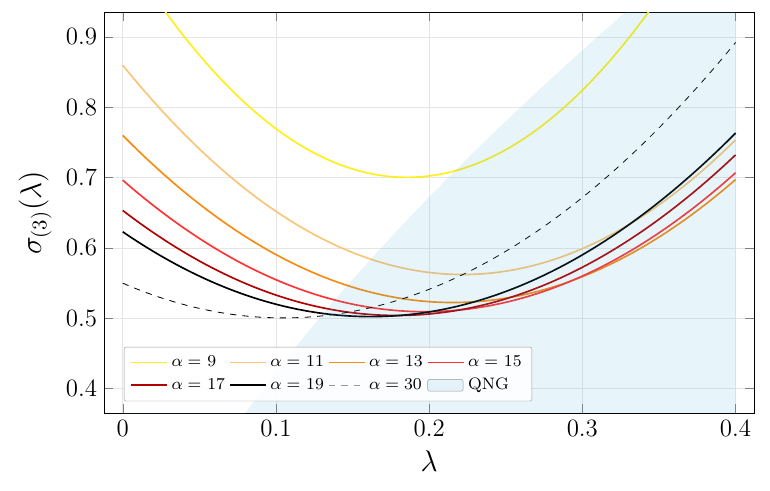}
		\caption{	\label{fig:s3-n-cubic}
			Nonlinear squeezing of the vibrational state with the exact numerical evolution given by the interaction potential in Eq. \eqref{eq:V}. Each curve corresponds to a different initial coherent amplitude, $\alpha$, at the time $\tau = \pi/|\omega_{1}|$. Here, the blue-shaded zone marked as QNG represents NLS values corresponding to quantum non-Gaussian states.
		}
	\end{figure}
	%.
    
    We devote this section to studying the non-Gaussian properties of our proposed vibrational CPS with the aid of the NLS, which also serves as a witness of the capability of the state for CV quantum computing \cite{Brauer2021,Kala2025}. One can motivate this quantity by considering that  the cubic phase gate, $e^{i\,\gamma\, \hat{x}^{3}}$, does not modify the position operator, however, its conjugate momentum operator $\hat{p} = i(\hat{a}^{\dag} -\hat{a})/\sqrt{2}$ is transformed into the following nonlinear quadrature
    \begin{equation}
    \label{eq:transp}
        e^{-i\,\lambda\, \hat{x}^{3}}\,
        \hat{p}\,
        e^{i\,\lambda\, \hat{x}^{3}}
        =
        \hat{p} + 3\,\lambda\,\hat{x}^{2},
    \end{equation}
    where we have introduced  $\lambda$ to distinguish it from $\gamma$; 
    the first parameterizes the quadrature, while the second parameterizes the CPS. Hence, it is natural to use this nonlinear quadrature as a prototype to quantify the non-Gaussian properties of an approximate CPS. We introduce the NLS as the variance of the nonlinear quadrature \cite{Miyata2016,Darren2022}, which is expressed as
    \begin{equation}
    \label{eq:def_s3}
		\sigma_{(3)} (\lambda)
		=
		\text{Var}\left(
		\hat{p} - 3\,\lambda\, \hat{x}^{2}
		\right),
	\end{equation}
	where the variance of an operator $\hat{A}$ is defined as Var$(\hat{A}) = \langle \hat{A}^{2} \rangle - \langle \hat{A} \rangle^{2}$ and $\lambda$ is a real parameter that is analogous to the quadrature angle used in linear squeezing. Using the NLS, we can characterize quantum states with non-Gaussian properties compatible with the CPS, as those states reduce the noise of the nonlinear quadrature below a certain threshold generated by any Gaussian state, which is given by the following expression
    \begin{eqnarray}
    \label{eq:gauss}
		\sigma_{G}
		&=&
		\left( \dfrac{3}{2} \right)^{5/3}
		| \lambda |^{2/3}.
	\end{eqnarray}
    This threshold is derived as the envelope curve of NLS for a vacuum state with linear squeezing \cite{Darren2022}, since the squeezed vacuum is the Gaussian state that minimizes Eq. \eqref{eq:def_s3}.
    
    We use the NLS to evaluate the properties of the vibrational state with the time-evolution generated by the interaction potential in Eq. \eqref{eq:V}. For the numerical evaluation, we used the expanded version of Eq. \eqref{eq:def_s3} in terms of the parameter $\lambda$, 
    \begin{align}
    \label{eq:coeff}
        \sigma_{(3)} (\lambda)
        &=
        C_{0} -3\,\lambda\,C_{1} +(3\,\lambda)^{2}\,  C_{2},
        \\
		C_{0}
		&=
		\text{Var}(\hat{p}),
        \;
		C_{1}
        =
		\text{Cov}(\hat{p},\, \hat{x}^{2}),
        \;
		C_{2}
        =
		\text{Var}(\hat{x}^{2}),\nonumber
    \end{align}
    with the covariance between operators $\hat{A}$ and $\hat{B}$ defined as Cov$(\hat{A},\, \hat{B}) = \langle \hat{A}\hat{B} + \hat{B}\hat{A} \rangle -2\, \langle \hat{A} \rangle \langle \hat{B} \rangle $. These expressions simplify the numerical evaluation of the NLS rather than directly using its definition in Eq. \eqref{eq:def_s3}.
	Therefore, the NLS is our figure of merit to discern wether a quantum state is compatible with a CPS, regardless of its non-Gaussian properties; for example, the Fock states, which are non-Gaussian, but have no compatibility with the CPS.
	
	%.
	\begin{figure}[tbp!]
		\centering
		\includegraphics[width=0.95\linewidth]{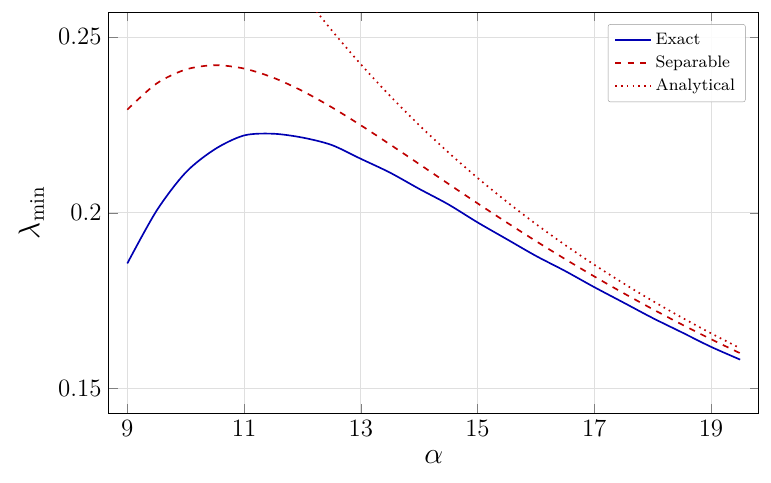}
		\caption{	\label{fig:min}
			Values of $\lambda_{\text{min}}$, according to Eq. \eqref{eq:l_min}, as a function of the initial coherent amplitude, $\alpha$. The solid lines correspond to the exact numerical evolution determined by the interaction potential in Eq. \eqref{eq:V}, the dashed line for the separable solution in Eq. \eqref{eq:evol_sep} and the dotted line shows the approximate analytical expression given in Eq. \eqref{eq:gamma}. The initial state is given in Eq. \eqref{eq:psi0} with the plus sign.
		}
	\end{figure}
	%.
    
    Now that we have introduced the explicit form of the CPS and the importance of the NLS for our analysis, we can present the form in which we evaluate the non-Gaussian properties of our generated vibrational state. In our case, it is convenient to evaluate the NLS in the displaced picture given by $\ket{\phi} = \hat{D}(\alpha)\, \ket{\psi(\tau)}$, Eq. \eqref{eq:psi_dis}. In this frame, the nonlinear quadrature is also transformed and therefore, the coefficients in Eq. \eqref{eq:coeff} take the following form
		\begin{align}
        \label{eq:coeff2}
			C_{0}
			&=
			\text{Var} (\hat{p})_{{\psi}},
			\nonumber\\
			C_{1}
			&=
			\text{Cov} (\hat{p},\, \hat{x}^{2})_{{\psi}}
			+
			\sqrt{8}\, r\,
			\text{Cov} (\hat{p},\, \hat{x})_{{\psi}},
			\\
			C_{2}
			&=
			\text{Var} (\hat{x}^{2})_{{\psi}}
			+
			\sqrt{8}\, r\,
			\text{Cov} (\hat{x}^{2},\, \hat{x})_{{\psi}}+
			8\, r^{2}\,
			\text{Var} (\hat{x})_{{\psi}},\nonumber
		\end{align}
	where we have used the subscript $\psi$ to denote that the variances and covariances are evaluated with the vibrational state $\ket{\psi(t = \tau)}$ given in Eq. \eqref{eq:psi_vib2}.

    In Fig. \ref{fig:s3-n-cubic}, we show the numerical NLS for the fixed time $\tau = \pi/|\omega_{1}|$ as in the sketch in Fig. \ref{fig:evol_cohe}. Two remarkable features are observed. First, the non-Gaussian vibrational features allow us to get values of $\sigma_{(3)}$ below the Gaussian threshold, which is a clear signature of its compatibility with a CPS. The second feature shows that for larger values of the initial coherent amplitude, the behavior of the evolved vibrational state is closer to that obtained for an ideal CPS, Eq. \eqref{eq:CPS}, with the cubicity value given by Eq. \eqref{eq:gamma}, that is $\ket{ \gamma = \tilde{\gamma} }$. To demonstrate it, we use the value of $\lambda$ at which the NLS has its minimum, namely $\lambda_{\text{min}}$. Using the expanded form of $\sigma_{(3)}$ in Eq. \eqref{eq:coeff}, we obtain
	\begin{eqnarray}
    \label{eq:l_min}
		\lambda_{\text{min}}
		&=&
		\dfrac{C_{1}}
		{6\, C_{2}},
	\end{eqnarray}
	where the coefficients $C_{1}$ and $C_{2}$ are given in its expanded form in Eq. \ref{eq:coeff2}.

    In Fig. \ref{fig:min}, we compare the values of $\lambda_{\text{min}}$ as a function of the initial coherent amplitude obtained with the exact numerical evolution given by Eq. \eqref{eq:V}, our approximate separable solution in the displaced picture approach, Eq. \eqref{eq:mean_field}, and the analytical cubicity in Eq. \eqref{eq:gamma}. We observe that the three results become closer for larger initial $\alpha$ and, as we have shown with the fidelity analysis, the accuracy of our separable solution increases with it, which is a consequence of the assumption we made to get the approximate solution in Eq. \eqref{eq:evol_sep}. One should keep in mind, however, that excessively large values of $\alpha$ diminish the value of $\lambda_{\text{min}}$ and therefore the cubicity. For this reason, the presented approach is valid for moderately large values of $\alpha$.
    
	%%%%%%%%%%%%%%%%%%%%%%%%%%%%%%%%%%%%%%%%%%%%%%%%%%%%%%%%%%%%%%%%%%%
	\section{Conclusions} \label{sec:conclusions}
	%%%%%%%%%%%%%%%%%%%%%%%%%%%%%%%%%%%%%%%%%%%%%%%%%%%%%%%%%%%%%%%%%%%
	
	In this work, we have presented a method to generate quantum states in the vibrational mode of a trapped ion with features that make them a highly accurate approximation of the cubic phase state. We exploited the nonlinear interaction between the electronic and vibrational degrees of freedom in a trapped ion to generate a vibrational state with non-Gaussian properties.

    We derived an approximate separable solution for the time-dependent state of the trapped ion under the assumption of an initial coherent state with large occupation number for the vibrational mode. Our separable solution exhibits high fidelity when compared with the exact numerical evolution; this validates our approximation. That approach allows us to study the vibrational state separately and, consequently, we achieved a better understanding of its behavior. We visualize the non-Gaussian features of our generated vibrational state in phase space with the aid of the Wigner function. We found the distinctive pattern of oscillations with negative values characteristic of the cubic phase state, but with a clear global deformation due to the intensity dependence of the interaction potential in the nonlinear Jaynes-Cummings model. The high fidelity between our vibrational cubic phase state and an ideal one corroborates the accuracy of the approximation. As a final characterization of our generated vibrational state, we use the nonlinear squeezing to quantify its usefulness for continuous variable quantum computation; the resulting values demonstrate its strong performance as a non-Gaussian resource to approximate the cubic phase gate. Additionally, for very large values of the initial coherent amplitude of the vibrational state, we noticed that the coherent state is preserved under the linear evolution, consistent with our findings in Ref.~\cite{MauricioTorres2025}.
    
    In summary, our results prove that it is possible to generate an approximate cubic phase state in the vibrational mode of a trapped ion that can be used as a non-Gaussian resource in quantum circuits for quantum computation. Beyond trapped-ion systems, this analysis can be adapted to a bosonic system with a time-evolution governed by a nonlinear Hamiltonian with an intensity dependence.

    %%%%%%%%%%%%%%%%%%%%%%%%%%%%%%%%%%%%%%%%%%%%%%%%%%%%%%%%%%%%%%%%%%%
	\begin{acknowledgments}
        This work was supported by CONAHCYT (SECIHTI-Mexico) Research Grant CF-2023-I-1751. C.V.-V. was supported by a postdoctoral grant under the SECIHTI-Mexico program \textit{Estancias posdoctorales por México 2022(1)}.
    \end{acknowledgments}
	%%%%%%%%%%%%%%%%%%%%%%%%%%%%%%%%%%%%%%%%%%%%%%%%%%%%%%%%%%%%%%%%%%%
    %%%%%%%%%%%%%%%%%%%%%%%%%%%%%%%%%%%%%%%%%%%%%%%%%%%%%%%%%%%%%%%%%%%
    
    % Bibliography
    
	\bibliography{references.bib}
	\bibliographystyle{apsrev4-2}

\end{document}